\documentclass{article}
\usepackage{graphicx}
\usepackage{amsmath,amssymb}
\usepackage{graphics,color,array,calc,rotating,epsfig,psfrag}
\usepackage{hyperref}
\numberwithin{equation}{section}
\usepackage{cite}
\usepackage{bm}
\usepackage{dcolumn}
\usepackage{float}
\usepackage{subfigure}
\usepackage{amsfonts}
\usepackage[mathscr]{eucal}
\def\be{\begin{equation}} \def\ee{\end{equation}}
\def\bea{\begin{eqnarray}} \def\eea{\end{eqnarray}}

\newcommand\prt{\partial}

\newcommand{\nn}{\nonumber}
\newcommand{\RN}[1]{%
  \textup{\uppercase\expandafter{\romannumeral#1}}%
}
\begin{document}
\baselineskip 18pt%
\begin{titlepage}
\vspace*{1mm}%
\hfill%
\vspace*{15mm}%
\hfill
\vbox{
    \halign{#\hfil         \cr
%         hep-th/yymmnnn\cr
         IPM/P-2020/051  \cr
          } % end of \halign
      }  % end of \vbox
\vspace*{20mm}

\centerline{{\Large {\bf Holographic complexity for black branes with momentum relaxation}}}
\vspace*{5mm}
\begin{center}
{Davood Mahdavian Yekta$^{a}$, H. Babaei-Aghbolagh$^{b}$, Komeil Babaei Velni$^{c,d}$, H. Mohammadzadeh$^{b}$}\\
\vspace*{0.2cm}
{\it $^{a}$Department of Physics, Hakim Sabzevari University, P.O. Box 397, Sabzevar, Iran\\
$^{b}$Department of Physics, University of Mohaghegh Ardabili,
P.O. Box 179, Ardabil, Iran\\
$^{c}$Department of Physics, University of Guilan, P.O. Box 41335-1914, Rasht, Iran\\
$^{d}$School of Physics and School of Particles and Accelerators,\\
Institute for Research in Fundamental Sciences (IPM), P.O. Box 19395-5531, Tehran, Iran\\
}

 \vspace*{0.5cm}
{E-mails: {\tt d.mahdavian@hsu.ac.ir, h.babaei@uma.ac.ir, babaeivelni@guilan.ac.ir,  mohammadzadeh@uma.ac.ir}}
\vspace{1cm}
\end{center}

\begin{abstract}
We employ the ``complexity equals action'' conjecture to investigate the action growth rate for the charged and neutral AdS black branes of a holographic toy model consisting of Einstein-Maxwell theory in $d + 1$-dimensional bulk spacetime with $d - 1$ massless scalar fields which is called Einstein-Maxwell-Axion (EMA) theory. From the holographic point of view, the scalar fields source a spatially dependent field theory with momentum relaxation on the boundary, which is dual to the homogeneous and isotropic black branes. We find that the growth rate of the holographic complexity within the Wheeler-DeWitt (WDW) patch saturates the corresponding Lloyd's bound at the late time limit. Especially for the neutral AdS black branes, it will be shown that the complexity growth rate at late time vanishes for a particular value of relaxation parameter $\beta_{max}$ where the temperature of the black hole is minimal. Then, we investigate the transport properties of the holographic dual theory in the minimum temperature. A non-linear contribution of the axion field kinetic term in the context of k-essence model in the four-dimensional spacetime is considered as well. We also study the time evolution of the holographic complexity for the dyonic AdS black branes in this model.
\end{abstract}

\end{titlepage}

%%%%%%%%%%%%%%%%%%%%%%%%%%%%%%%%%%%
\section{Introduction}

The AdS/CFT correspondence \cite{Maldacena:1997re,Witten:1998qj,Gubser:1998bc}, as the most important realization of the holographic principle \cite{tHooft:1993dmi,Susskind:1994vu}, relates a gravity theory in an asymptotically anti-de Sitter (AdS) spacetime in the bulk to a conformal field theory (CFT) without gravity living on the boundary of this spacetime. It suggests the non-trivial connections between different areas of physics, in particular between general relativity and quantum information theory. One of the outstanding developments in this correspondence is the seminal work of Ryu and Takayanagi \cite{Ryu:2006bv,Ryu:2006ef}, which provides a holographic dictionary for the calculation of the entanglement entropy of the boundary theory. According to this proposal, the entanglement entropy of the boundary theory is equivalent to the area of a certain minimal surface in the bulk geometry. In other words, the dynamics of the bulk spacetime emerges from the quantum entanglement of the boundary theory \cite{Faulkner:2013ica}. However, the entanglement entropy may not be enough to probe the degrees of freedom in black holes interior since the volume of black holes continues growing even if spacetimes reach the thermal equilibrium \cite{Susskind:2014moa}. It is believed that the quantum complexity is the correct quantity which can continue to grow even after reaching the thermal equilibrium, similar to the growth of black hole interior.

In the framework of the quantum information theory, the quantum complexity is defined by the minimal number of quantum gates needed to build a target state from a reference state \cite{watrous2009quantum,Aaronson:2016vto}.  However, the AdS/CFT correspondence provides two proposals to compute the complexity of states in the boundary quantum field theory of the two-sided AdS black holes. The first one is the complexity=volume (CV) conjecture which assumes that the quantum complexity of the CFT on boundary is dual to the maximum volume of the Einstein-Rosen Bridge, i.e. $V$, in the bulk spacetime \cite{Susskind:2014rva,Stanford:2014jda},
\be\label{CV} \mathcal{C}_{V}\!\sim\! \frac{V}{G\ell_{AdS}},\ee
where $\ell_{AdS}$ is the radius of curvature of the AdS spacetime and G is the Newton's constant. The second is the complexity=action (CA) conjecture which states that the quantum complexity on boundary is associated to the gravitational action evaluated on a region of the Wheeler-DeWitt (WDW) patch in the bulk spacetime \cite{Brown:2015bva,Brown:2015lvg},
\be\label{CA} \mathcal{C}_A \!\sim \!\frac{I_{WDW}}{\pi\hbar}.\ee
Moreover, when the WDW patch besides space/time-like boundaries includes null boundary surfaces \cite{Hayward:1993my}, which can join with each other, the strategy in CA conjecture has suggested in Ref. \cite{Lehner:2016vdi}.

In general, the holographic complexity has been suggested in Ref. \cite{Maldacena:2001kr}  for the eternal two-sided AdS black holes in the gravity side. From the field theory point of view, this geometry is dual to a thermofield double state as following
\be\label{TFD} |\psi_{TFD}\rangle=\frac{1}{\sqrt{Z}}\sum_j e^{-{E_j}/({2T})} e^{-i E_{j}(t_{L}+t_{R})}|E_j\rangle_L |E_j\rangle_R\,,\ee
where $L$ and $R$ refer to the two copies of the
boundary CFTs. The entanglement between $L$ and $R$ copies is due to the Einstein-Rosen bridge that connects two regions. Since the complexity is conjectured to grow with time and this property is also shared with the Einstein-Rosen bridge, in Refs. \cite{Maldacena:2001kr,Maldacena:2013xja}, it was conjectured that the complexity could be identified with the volume of the maximal co-dimension one surface that ends to the boundary times $t_L$ and $t_R$.

The growth rate of the holographic complexity is one of the noticeable outcomes in the CA conjecture that asserts the late time growth rate is proportional to $2M/\pi$, independent of the boundary curvature and the spacetime dimensions \cite{Brown:2015bva,Brown:2015lvg}. It was also suggested that this quantity has an upper bound which is proportional to the total energy of the system
\be\label{LB} \dot{\mathcal{C}}\leq \frac{2E}{\pi\hbar},\ee
 where this inequality is known as the Lloyd's bound \cite{Lloyd} derived from the Margolus-Levitin theorem \cite{Margolus:1997ih} under the assumption that each gate will evolve from a generic state into an orthogonal state. In the gravitational picture, the mass of the black hole, $M$, is regarded as the energy, $E$. The generalization of this bound for the charged and rotating black holes are given in Refs. \cite{Brown:2015lvg,Cai:2016xho}, respectively as follows
 \be \label{CRB}
 \frac{dI_{WDW}}{dt}\leq2\left[(M-\mu Q)-(M-\mu Q)_{gs}\right],\qquad
 \frac{dI_{WDW}}{dt}\leq2\left[(M-\Omega J )-(M-\Omega J )_{gs}\right],
 \ee
where $\mu$ and $\Omega$ are the chemical potential and angular velocity of the black holes respectively, $Q$ and $J$ are the black hole charge and angular momentum, respectively. Intuitively, these conserved charges impose a tighter bound because they provide a barrier to the rapid complexification and consequently, some energy is tied up in non-computing degrees of freedom. The subscript ``gs'' denotes the ground state of the black hole. However, it is known \cite{Brown:2015bva,Lloyd} that this proposal is violated at least at early times in holographic theories \cite{Carmi:2017jqz,Kim:2017qrq,Swingle:2017zcd,An:2018xhv} and in
sufficiently exotic computational setups in non-holographic theories as well \cite{Jordan:2017vqh,Cottrell:2017ayj,Deffner:2017cxz}. The late time violation of this bound has been considered for holographic models in Refs.~ \cite{Swingle:2017zcd,An:2018xhv,Brown:2017jil,Couch:2017yil,Moosa:2017yiz,Mahapatra:2018gig,An:2018dbz}.
Different aspects of the holographic complexity such as the subregion complexity \cite{Alishahiha:2015rta,Carmi:2016wjl,Ben-Ami:2016qex,Alishahiha:2018lfv}, the UV divergencies of complexity \cite{Cai:2016xho,Chapman:2016hwi,Reynolds:2016rvl,Kim:2017lrw,Alishahiha:2019cib}, higher derivative gravities \cite{Alishahiha:2017hwg,Cano:2018aqi,Jiang:2018pfk,An:2018dbz,Jiang:2018sqj,Jiang:2019kks,Ghodsi:2020qqb}, and the Einstein-Maxwell-dilaton gravity \cite{Swingle:2017zcd,An:2018xhv,HosseiniMansoori:2017tsm,Cai:2017sjv} have been studied in both of the CV and CA conjectures.
The attempts to define the complexity more rigorously in the quantum field theory and in a continuous way, where interestingly their results in different setups match with results from the holography could be found in Refs.~ \cite{Chapman:2017rqy,Jefferson:2017sdb,Czech:2017ryf,Belin:2018bpg,Chapman:2018hou,Khan:2018rzm,Camargo:2019isp}.

The holographic correspondence has also provided us a powerful tool to study the behavior of the strongly correlated materials in the condensed matter (CM) physics \cite{Hartnoll:2009sz,Herzog:2009xv,Sachdev:2011wg} which can be mapped to the classical bulk gravity. Especially, much attention has
been paid to the holographic description of systems with the momentum relaxation. Such systems with broken translational symmetry are needed to give a realistic description of materials in many CM systems \cite{Liu:2012tr,Ling:2013aya,Ling:2014bda,Mozaffara:2016iwm,Cremonini:2018xgj,Cremonini:2019fzz}. Since momentum is conserved in a system with translational symmetry, a constant
electric field can generate a charge current without current dissipation in the presence of
a non-zero charge density. Thus, the conductivity of the system would become divergent
at the zero frequency. In more realistic CM materials, the momentum is not
conserved due to impurities or a lattice structure, leading to a finite DC conductivity.
There are various ways to achieve momentum dissipation,
such as periodic potentials, lattices and breaking diffeomorphism invariance \cite{Horowitz:2012ky,Davison:2013jba,Blake:2013bqa,Ling:2013nxa,Donos:2013eha,Andrade:2013gsa,Donos:2014yya,Davison:2015bea}.
However, there are two well-known strategies to produce momentum dissipation by inclusion matter fields that breaks the translational invariance in the dual field theory. The case of scalar fields (EMA theory) that linearly depends on the horizon coordinates as given in Ref. \cite{Andrade:2013gsa}, and the case of massive gravity theories which presents a broken diffeomorphism invariance in the bulk as done in Ref. \cite{deRham:2010kj}.

As the main purpose of this paper, we employ the CA conjecture to study the holographic complexity and its time evolution in the EMA theory with momentum relaxation by following the approach used in Ref. \cite{Carmi:2017jqz}. In particular, we compute these quantities for the charged and neutral AdS$_{d+1}$ black branes and investigate the Lloyd's bound for these solutions. Though one can also employ the CV conjecture to study the evolution of the holographic complexity, but there are some unsatisfactory elements that are more interested in CA. For instance, in CV picture, we need to introduce an arbitrary length scale by hand while in CA it is not necessary, or in CV picture one should find the volume of a maximal slice in the bulk while CA associates with the boundary state on the entire WDW patch and is easier to work with than a special maximal volume. The CA conjecture can also satisfy the Lloyd’s complexity growth bound in very general cases \cite{Brown:2015bva,Brown:2015lvg}. In fact, CA inherits all the nice features of CV duality and none of the unsatisfactory elements.

We provide an analytical discussion for the effects of the strength of momentum relaxation on the desired quantities. The results show that the Lloyd's bound is saturated only at late time limit and there is an upper bound on the strength of the momentum relaxation in each sector which provides a minimum temperature for the gravitational system to has positive energy. This specific value of the relaxation parameter also motivates us to investigate the characteristic properties of the strongly correlated materials in CM physics \cite{Hartnoll:2014lpa}. In fact, we study thermal conductivity and diffusivity of strongly coupled theories which are holographically dual to the EMA theory \cite{Maldacena:2015waa,Shenker:2013pqa,Roberts:2014isa}.
Inspired the strange metals characterized by a minimum Planckian relaxation timescale $\tau_L$, it has been proposed in Ref.~\cite{Hartnoll:2014lpa} that there is a universal bound for the diffusivities in the incoherent limit $\mathcal{D}_T\geq v_B^2 \tau_L$
where $v_B$ is a characteristic velocity of the so-called butterfly velocity \cite{Shenker:2013pqa}.
We will show that the diffusion constant in the EMA theory saturates this bound in the corresponding minimum temperature.
In addition, we examine the effect of the non-linear contribution of the scalar field kinetic term \cite{ArmendarizPicon:1999rj,Baggioli:2014roa,Cisterna:2017jmv} on the complexity growth rate in the four-dimensional spacetime. This theory is known as the k-essence model of dark energy \cite{ArmendarizPicon:1999rj} in which the acceleration of the Universe (both at early and late times) can be driven by the kinetic energy instead of the potential energy of the scalar field. The time evolution of the holographic entanglement entropy and complexity under a thermal quench has been recently studied for EMA theory in Ref. \cite{Zhou:2019xzc}, in the context of the CV conjecture.

The structure of this paper is organized as follows: in section 2, we review the EMA theory with momentum relaxation and study the time-evolution of the holographic complexity for the charged/neutral AdS$_{d+1}$ black branes. In the context of CA conjecture, we consider the WDW patch that includes null sheets bounding the bulk and joint terms, and investigate how the holographic complexity approaches the late time limit on them. We will also investigate the transport properties of the dual theory in CM physics from the holographic point of view. In section 3, we study the dyonic AdS black branes in the presence of the non-linear kinetic term in the k-essence model. In calculating the growth rate, the contribution of the Maxwell surface term to the action will be considered, as well. Finally, the section 4 is dedicated to a brief summary and concluding remarks.
%@@@@@@@@@@@@@@@@@@@
\section{EMA theory with momentum relaxation}
In order to have momentum relaxation and finite conductivity, it is essential to construct holographic models with broken translational symmetry. Thus, we consider a  model in which the Einstein-Maxwell action in the $(d+1)$-dimensional spacetime is supplemented by $d-1$ massless scalar fields that break the translational invariance of the boundary theory in the context of the AdS/CFT duality \cite{Andrade:2013gsa}. This theory is known as the EMA theory in Horndeski theories of modified gravity \cite{Horndeski:1974wa,Kobayashi:2011nu}. The scalar axion fields enter the bulk action only through the kinetic term $\prt_{\mu}\psi_{I} $ and the sources are linear in the boundary, i.e. $\psi^{(0)}_{I} \propto \beta_{I i} x^i$, where $\beta$ represents the strength of the momentum relaxation.

The action of this holographic model in the bulk is described by \cite{Andrade:2013gsa}
\be\label{act1} I_{bulk}=\frac{1}{16\pi G}\int_{M} d^{d+1}x \,\sqrt{-g}\left[R-2\Lambda-\frac14 F_{\mu\nu}F^{\mu\nu}-\frac12 \sum_{I}^{d-1} (\prt \psi_{I})^2\right],\ee
where $G$ is a $(d+1)$-dimensional gravitational constant and $\Lambda\!=\!-d(d-1)/2L^2$  is a cosmological term. The action includes the field strength $F_{\mu\nu}\!=\!\prt_{\mu}A_{\nu}-\prt_{\nu}A_{\mu}$ of a $U(1)$ gauge field $A_{\mu}$ and $d-1$ massless scalar fields $\psi_{I}$.
The model admits the homogeneous and isotropic charged AdS$_{d+1}$ black brane solutions of radius $L$ with the non-trivial scalar field sources. They are described by the following ansatz
\be \label{ansatz1} ds^2 = - f(r) dt^2 + \frac{dr^2}{f(r)} + r^2 \delta_{i j} dx^i dx^j,  \qquad A=A_{t}(r) dt, \qquad \psi_I = \beta_{I i}\, x^i,\ee
where $i$ labels the $d-1$ spatial $x_{i}$ directions and $I$ is an internal index that labels the $d-1$ scalar fields. Substituting the ansatz (\ref{ansatz1}) in the equations of motion derived from the action (\ref{act1}), we find that
\be\label{sol1} f(r)=r^2-\frac{\beta^2}{2(d-2)}-\frac{m_{0}}{r^{d-2}}+\frac{q^{2}}{r^{2(d-2)}}, \qquad A_t (r)=\sqrt{\frac{2\,(d-1)}{d-2}}\,q\left(\frac{1}{r_{h}^{d-2}}-\frac{1}{r^{d-2}}\right),\ee
where
\be \beta^2\equiv\frac{1}{d-1}\sum_{i}^{d-1} \vec{\beta}_{i}\cdot \vec{\beta}_{i}, \qquad \vec{\beta}_{i}\cdot \vec{\beta}_{j}=\sum_{I}{\beta}_{Ii}\, {\beta}_{I j} =\beta^2 \delta_{i j} \qquad \forall \, i, j.\ee
 Note that for the AdS radius we set $L\!=\!1$ in the rest of the paper.

 The mass parameter $m_0$, which is proportional to the energy density of the brane, is computed from $f(r_h)\!=\!0$ as follows
\be\label{mass1} m_0 = r_h^d \left(1+ \frac{q^2}{r_h^{2(d-1)}}- \frac{1}{2(d-2)}\frac{\beta^2}{r_h^2} \right),\ee
where $r_h$ is the location of the event horizon. This is related to the mass of the brane with \cite{Myers:1999psa}
\be\label{mass2} M=\frac{(d-1) V_{d-1}}{16\pi G}m_0.\ee
Here, $V_{d-1}$ is the dimensionless volume of the relevant spatial geometry.
The parameter $q$ is related to the charge of the brane through the Gauss's law with
\be\label{CH}  Q =\frac{\sqrt{2(d-1)(d-2)}\, V_{d-1}}{16\pi G }
\,q\,.\ee
The Hawking temperature and the entropy of the branes are given by
\be \label{TS1} T= \frac{f^\prime (r_h)}{4 \pi}=\frac{1}{4\pi}\left(d\,r_+ -\frac{\beta^2}{2r_h} -\frac{(d-2) q^2}{r_h^{d-1}}\right),\qquad  S=\frac{V_{d-1}}{16\pi G} 4\pi r_h^{(d-1)}.\ee
Since the blackening factor in (\ref{sol1}) has two real roots $r_+$ and $r_-$ (where $r_+> r_-$), corresponding to the outer and inner horizons in which $f(r_{+})=f(r_{-})=0$, we can define a chemical potential for both of them as follows
\be\label{CP} \mu_+=\frac{\prt M}{\prt Q}\Bigg|_{V,S}=\sqrt{\frac{2(d-1)}{d-2}}\,\frac{q}{r_+^{d-2}}\,,\qquad \mu_-=\frac{\prt M}{\prt Q}\Bigg|_{V,S}=\sqrt{\frac{2(d-1)}{d-2}}\,\frac{q}{r_-^{d-2}}\,.\ee
 Various features of thermodynamics of this solution have been extensively studied in Ref. \cite{Fang:2017nse}.
%@@@@@@@@@@@@@@@@@@@@@@
\subsection{Complexity of charged black branes via CA conjecture }
We use the CA conjecture (\ref{CA}) to compute the holographic complexity for the charged AdS$_{d+1}$ black branes in the EMA theory. The essential ingredient in this method is to evaluate the action on a WDW patch \cite{Brown:2015bva,Brown:2015lvg}. However, we follow the method of Ref. \cite{Carmi:2017jqz} in which not only the action in the WDW patch includes the bulk theory and the Gibbons-Hawking-York (GHY) boundary term \cite{York:1972sj,Gibbons:1976ue}, but also embraces boundary segments of joint terms due to the intersection of the time-like, space-like, and null boundaries \cite{Hayward:1993my}. This will be the general strategy that would be followed in the rest of the paper. The contribution of the GHY surface terms are
\be\label{surf} I_{bdy}=\frac{1}{8\pi G}\int_{\mathcal{B}} d^{d} x \sqrt{\gamma} K-\frac{1}{8\pi G}\int_{\mathcal{B'}} d\lambda d^{d-1}\theta \sqrt{\gamma}\,\kappa,\ee
where $K$ is the trace of the extrinsic curvature $K_{\mu\nu}=-\gamma_{\mu}^{\rho} \gamma_{\nu}^{\sigma} \nabla_{(\rho}n_{\sigma)}$, $\gamma_{\mu\nu}$ is the induced metric on the boundary and $n_{\mu}$ is the outward pointing unit normal vector to the boundary. $\kappa$ is the surface term for the null segments which measures the failure of the null generators to be affinely parametrized which is assumed to vanish, i.e., does not have any contribution to CA for null segments.

The joint actions are given by
\be\label{joint} I_{joint}=\frac{1}{8\pi G}\int_{\Sigma}  d^{d-1}x \sqrt{\sigma}\,\varrho+ \frac{1}{8\pi G}\int_{\Sigma '} d^{d-1}x \sqrt{\sigma}\,a,\ee
in which $\varrho$ appears when we have the intersection of time-like or space-like boundaries, the so called Hayward terms \cite{Hayward:1993my}, while $a$ is required when one or both of the intersecting boundaries belong to null surfaces \cite{Lehner:2016vdi}.
The general rules for the construction of the former joint terms could also be found in Refs.\cite{Lehner:2016vdi,Carmi:2016wjl}. In particular, for time-like normals ${\bf{t}}_i$, space-like normals ${\bf{n}}_i$ and auxiliary unit vectors $\hat{\bf{t}}_i$ and $\hat{\bf{n}}_i$, $\varrho$ is given by
\bea\label{tsj}
\varrho&\!\!\!=\!\!\!&arccosh |{\bf{t}}_1\cdot {\bf{t}}_2| \qquad sign(\varrho)=-sign({\bf{t}}_1\cdot {\bf{t}}_2)\,sign(\hat{\bf{n}}_1\cdot {\bf{t}}_2),\nn\\
\varrho&\!\!\!=\!\!\!&arccosh |{\bf{n}}_1\cdot {\bf{n}}_2| \qquad sign(\varrho)=-sign({\bf{n}}_1\cdot {\bf{n}}_2)\,sign({\bf{n}}_1\cdot \hat{\bf{t}}_2),\\
\varrho&\!\!\!=\!\!\!&arcsinh|\epsilon {\bf{t}}_1\cdot {\bf{n}}_2| \qquad \epsilon =-sign({\bf{n}}_2\cdot \hat{\bf{n}}_1).\nn
\eea
However, these are not relevant here since all of the joints that we consider in the WDW patches involve at least one null surface. The latter ones are also defined in the next subsection appropriately. There is also a counterterm action for the null surfaces as
\be\label{cterm} I_{ct}=\frac{1}{8\pi G} \int_{\mathcal{B'}} d\lambda \,d^{d-1}\theta \sqrt{\gamma}\, \Theta \log{(\ell_{c}\Theta)}\,,\ee
 which is introduced to ensure reparametrization invariance on the null boundaries. $\Theta$ is the expansion parameter that is related to the induced metric as
\be\label{amb}\Theta=\prt_{\lambda} \log{\sqrt{\gamma}},\ee
and $\ell_{c}$ is an arbitrary length scale. The precise definition of parameters, boundary metrics, and comprehensive discussions of these actions are given in Ref. \cite{Lehner:2016vdi}. In brief, the total action is defined by
 \be\label{tact} I_{tot}=I_{bulk}+I_{bdy}+I_{joint}+I_{ct}\,.\ee
Of course, there may be a boundary term for the Maxwell field in this action that does not change the equations of motion, but it affects the variational principle for the Maxwell field and one should change the boundary conditions consistently \cite{Goto:2018iay}. We will consider the contribution of such a term for charged geometries in the next section.

Due to the presence of null boundaries in the total action, it is more convenient to introduce the ingoing and outgoing coordinates
\be \label{efc} v=t+r^{*} (r)\,,\qquad u=t-r^{*} (r),\ee
where $r^{*}$ is a tortoise coordinate defined as
\be \label{tc} r^{*} (r) =  \int  \frac{d r}{f ( r)} \,,\ee
with asymptotic behavior
\be \label{tca} \lim_{r \rightarrow \infty} r_{}^{*} (r) = {r}_{\infty}^{*}.\ee

\begin{figure}[H]
\centering
\includegraphics[width=8cm,height=10cm]{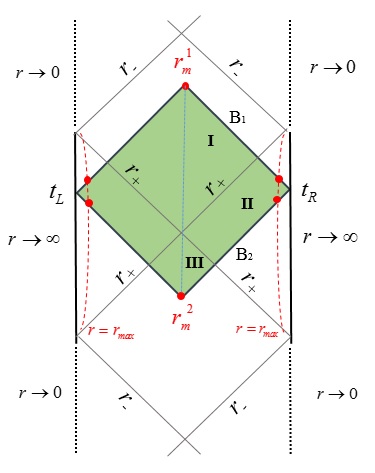}
\caption{Penrose diagram of the WDW patch for charged AdS black brane in symmetric configuration ($t_L=t_R$). $r\to 0$ is the singular surface and $r\to \infty$ is the asymptotic boundary surface. The red dashed lines correspond to UV cutoff surfaces at $r=r_{max}$ and $r_m^{1}, r_m^{2}$ are the meeting points of null boundaries in the bulk.}
\label{f1}
\end{figure}

{\bf{\emph{WDW patch:}}} In order to study the evolution of complexity for the action (\ref{tact}), we draw the Penrose diagram of causal structure of the charged AdS black brane described by (\ref{sol1}) in Fig.~(\ref{f1}). Following \cite{Carmi:2017jqz}, the corresponding WDW patch is denoted by the shaded region which is bounded by the light sheets sent from the two asymptotic time slices $t_L$ and $t_R$. Without lose of generality, we choose the symmetric configuration for the time slices, i.e. $t_L=t_R\equiv t/2$. In the next subsection, we evaluate the gravitational action on this patch as the boundary time increases.

The patch includes two UV cutoff surfaces near the asymptotic boundary regions at $r=r_{max}$ which are denoted by red dashed lines in Fig.~(\ref{f1}). In fact, the null boundaries of the WDW patch begin from the UV cutoff surface at $r = r_{max}$ and go through the bulk spacetime. There are two meeting points in the bulk which come from the intersecting future boundaries at $r = r_{m}^1$ and past boundaries at $r = r_{m}^2$. The time evolution of the WDW patch can be encoded in the time dependence of these points. These satisfy the following relations
\be\label{r1r2} \frac{t}{2}+{r}_{\infty}^{*}-r_{}^{*} (r_{m}^1)=0,\qquad \frac{t}{2}-{r}_{\infty}^{*}+r_{}^{*} (r_{m}^2)=0,\ee
in which by using (\ref{tc}) their time evolution is given by
\be\label{tr1r2} \frac{dr_m^{1}}{dt}=\frac{f(r_m^1)}{2},\qquad \frac{dr_m^2}{dt}=-\frac{f(r_m^2)}{2}.\ee
The null boundaries of the right sector of the corresponding WDW patch are
\be\label{BB} B_1: \frac{t}{2}=r_{}^{*} (r)-{r}_{\infty}^{*}\,,\qquad B_2: -\frac{t}{2}=r_{}^{*} (r)-{r}_{\infty}^{*}\,.\ee
These equations are important in the study of time evolution of the total action (\ref{tact}).

From the holographic point of view \cite{Hartnoll:2009sz}, it has been proposed in Ref. \cite{Carmi:2017jqz} that this black hole geometry with $U(1)$ symmetry is dual to a charged thermofield double state
\be\label{CTFD} |\psi_{CTFD}\rangle=\frac{1}{\sqrt{Z}}\sum_{j,l} e^{-{E_j-\mu Q_l}/({2T})} e^{-i E_{j}(t_{L}+t_{R})}|E_j,-Q_l\rangle_L |E_j,Q_l\rangle_R\,,\ee
where, comparing with state (\ref{TFD}), in addition to the temperature $T$, this state has a chemical potential $\mu$ and electric charge $Q$. The density matrix of the corresponding grand canonical ensemble characterized by $T$ and $\mu$ is obtained by tracing out the states in its boundary.
%@@@@@@@@@@@@@@@@
\subsubsection{The growth rate of complexity}
In the following we compute the growth rate of the holographic complexity on the WDW patch associated with a charged AdS black brane - see Fig.~(\ref{f1}).  In this regard, we consider the time dependence of the total action in (\ref{tact}). In the symmetric configuration of the WDW patch, we can perform the calculations only for the right side of the Penrose diagram or regions $\RN{1}$, $\RN{2}$, and $\RN{3}$ as depicted in Fig.~(\ref{f1}), and then multiply the result by a factor of two.
 %$$$$$$$$$$$$$$$$$$$$$$$$$

{\bf{\emph{Bulk action:}}} The Ricci scalar tensor and the kinetic terms of axion fields in this background are given by
\be\label{rvsv} R=-d(d+1)+\frac{(d-1)\beta^2}{2r^2}\,,\qquad -\frac12 \sum_{I}^{d-1} (\prt \psi_{I})^2=-\frac{(d-1)\beta^2}{2r^2},\ee
where the contribution of $\beta^2$ term in the scalar action is canceled by its contribution from the Einstein-Hilbert action, thus from the action (\ref{act1}) we have
\bea \label{bulk1} I_{bulk}&\!\!\!\!\!=\!\!\!\!\!&2\,\, \left(I_{bulk}^{\RN{1}}+I_{bulk}^{\RN{2}}+I_{bulk}^{\RN{3}}\right)\nn\\
&\!\!\!\!\!=\!\!\!\!\!&\frac{V_{d-1}}{8\pi G} \left[\int_{r_m^1}^{r_+}\left(\frac{t}{2}+{r}_{\infty}^{*}-r_{}^{*} (r)\right)+2\int_{r_+}^{r_{max}}\left({r}_{\infty}^{*}-r_{}^{*} (r)\right)+\int_{r_m^2}^{r_+}\left(-\frac{t}{2}+{r}_{\infty}^{*}-r_{}^{*} (r)\right) \right]\, I(r)\, dr,\nn\\
&\!\!\!\!\!=\!\!\!\!\!& I_{bulk}^0+I_{bulk}(t),
\eea
 where the integrand $I(r)$ is
 \be\label{bulkint} I(r)=r^{d-1} \left[-2 d + \frac{( d-2)\, q^2}{ 2\, r^{2(d-1) }} \right],\ee
and $I_{bulk}^0$ is the time independent part of the bulk action and only the first and third terms depend on the time through the Eqs.~(\ref{r1r2}).
 %$$$$$$$$$$$$$$$$$$$$$$$$$

 {\bf{\emph{Boundary surface action:}}} If we choose affinely parametrization for the null normals then the null surface term vanishes ($\kappa=0$), thus we only need to consider the GHY term coming from the surface at UV cutoff on the right side of the WDW patch. The trace of the extrinsic curvature for ansatz (\ref{ansatz1}) is given by
 \be \label{EX} K=\frac{1}{2}\left(\frac{\prt_r f(r)}{\sqrt{f(r)}}+\frac{2(d-1)}{r} \sqrt{f(r)}\right). \ee
 Following \cite{Carmi:2017jqz}, we define future directed normal vectors to evaluate $K$,
 \be \label{NV} r={\epsilon}:\quad {\bf t}=t_{\mu} dx^{\mu}=-\frac{dr}{\sqrt{-f(\epsilon)}},\qquad \qquad r=r_{max}:\quad {\bf s}=s_{\mu} dx^{\mu}=\frac{dr}{\sqrt{f(r_{max})}}.\ee
 Therefore, we obtain the contribution of surface action in Eq.~(\ref{surf}) for the charged AdS solution (\ref{sol1}) as
 \be\label{surf2} I_{bdy}=2\,I_{bdy}^{r=r_{max}}
 =\frac{r^{d-1}\,V_{d-1}}{8\pi G}\left(\prt_r f(r) +\frac{2(d-1)}{r} f(r)\right)({r}_{\infty}^{*}-r_{}^{*} (r))\Big|_{r=r_{max}}.
 \ee
As is obvious, the cutoff term at $r = r_{max}$ is independent of the time, so the boundary term has no contribution to the time evolution of the holographic complexity.
  %$$$$$$$$$$$$$$$$$$$$$$$$$

  {\bf{\emph{Joint action:}}} According to the WDW patch in Fig.~(\ref{f1}), there are different joint contributions at the intersection of null boundaries with the surfaces at $r=r_{max}$ and with each other at $r_{m}^1$ and $r_m^2$. It has been shown in Ref. \cite{Chapman:2016hwi} that the null joint contributions at the UV cutoff surfaces have no time dependence, so we need only to consider the last two joining points. Assume that $k_1$ and $k_2$ are the null vectors associated with two past null boundaries intersecting at $r_{m}^2$ which are defined by
  \be\label{kk} K_1=\xi\left(-dt+\frac{dr}{f(r)}\right), \qquad K_2=\xi\left(dt+\frac{dr}{f(r)}\right),\ee
 where $\xi$ is a normalization constant for null vectors. Those for null vectors associated with two future null boundaries (intersecting at $r_{m}^1$), i.e. $\tilde{k}_1$ and $\tilde{k}_2$, are similar. Following \cite{Lehner:2016vdi}, the joint term is defined by $a=\ln|-\frac12 k_1\cdot k_2|$ for the first set and $\tilde{a}=\ln|-\frac12 \tilde{k}_1\cdot \tilde{k}_2|$ for the second set, then the joint action (\ref{joint}) can be evaluated as
 \be\label{joint2} I_{joint}=-\frac{V_{d-1}}{8\pi G} \left[(r_{m}^1)^{d-1}\,\log\frac{|f(r_{m}^1)|}{\xi^2}+(r_{m}^2)^{d-1}\,\log\frac{|f(r_{m}^2)|}{\xi^2}\right],\ee
 where the time dependence of this contribution comes from Eqs. (\ref{tr1r2}).
 %$$$$$$$$$$$$$$$$$$$$$$$$$

 {\bf{\emph{Counterterm action:}}} In order to remove the ambiguity associated with the normalization of the null vectors, we need to add this boundary term to the action. Thus, we define the affine parameter $\lambda=r/\xi$ such that the total action with the counterterm does not depend on the parametrization of the null surfaces. In this parametrization, the expansion (\ref{amb}) takes the form \cite{Carmi:2017jqz}
 \be\label{amb2} \Theta=\frac{(d-1)\xi}{r}.\ee
 Therefore, the counterterm action in (\ref{cterm}) becomes
  \be\label{cterm2} I_{ct}=2\,(I_{ct}^{future}+I_{ct}^{past})=\frac{( d-1)\,V_{d-1}}{4\pi G} \left(\int_{r_m^1}^{r_{max}} r^{d-2} \log\frac{(d-1) \ell_c \xi}{r}\,dr+ \int_{r_m^2}^{r_{max}} r^{d-2} \log\frac{(d-1) \ell_c \xi}{r}\,dr\right).\ee
Again this contribution depends on the time through Eqs.~(\ref{tr1r2}).

 %$$$$$$$$$$$$$$$$$$$$$$$$$
\subsubsection{The late time behavior}
Now we can determine the rate of change of the holographic complexity by considering all of the above contributions
\be\label{CCAR1} \frac{d \mathcal{C}_{A}}{dt}=\frac{d}{dt} \left(\delta I_{bulk}+I_{joint}+I_{ct}\right),\ee
where $\delta I_{bulk}\equiv I_{bulk}-I_{bulk}^0$. Henceforth we will set $\pi \hbar=1$ in the CA conjecture in Eq.~(\ref{CA}) for simplicity.
Thus, the growth rate of the holographic complexity yields
\be\label{CCAR2} \frac{d \mathcal{C}_{A}}{dt}=\frac{ (d-1)\,V_{d-1}}{16\pi G}
\left[ \frac{2 q^2}{r^{d-2}} -\, r^{d-2} f(r) \log\frac{( d-1)^2 \ell_c^2 |f(r)|}{r^2}\right]_{r_m^2}^{r_m^1}\,.\ee
At late times, the future (past) corner approaches the inner (outer) horizon, such that due to the conditions $f(r_{+})=f(r_{-})=0$, the second term vanishes. This leaves the result
\be\label{CCAR3}\frac{d \mathcal{C}_{A}}{dt}\Bigg|_{t\to \infty}=\frac{ (d-1)\,V_{d-1}}{16\pi G} \left[\frac{2 q^2}{r^{d-2}}\right]_{r=r_+}^{r=r_-}=(M-\mu_+ Q)-(M-\mu_- Q),\ee
 where we have substituted from (\ref{CH}) and (\ref{CP}). The results in this limit are consistent with the calculations in Refs. \cite{Brown:2015lvg,Cai:2016xho} for the charged black holes without considering the joint and the counterterm actions directly.

For the complexity growth rate of the charged AdS black brane obtained in Eq.~(\ref{CCAR3}), it seems that in the $Q\to 0$ limit it vanishes, but as we know from general charged black holes, the $Q\to 0$ limit corresponds to $r_-\to0$, so we have $\mu_- Q\to 2M$ while $\mu_+ Q\to 0$. Therefore, in this limit we recover the case of neutral AdS branes for the Lloyd's bound, i.e. $d \mathcal{C}_A /dt=2M$.

%@@@@@@@@@@@@@@@@@@
\subsection{Complexity of neutral AdS black branes}
For the neutral black branes, it is sufficient to insert $q=0$ in the ansatz (\ref{sol1}). Therefore, the mass parameter $m_0$, which is proportional to the energy density of the brane, is computed from $f(r_h)\!=\!0$ that $r_h$ is the position of the event horizon,
\be\label{mass3} m_0=r_{h}^d\left(1- \frac{1}{2(d-2)}\frac{\beta^2}{r_h^2}\right),\ee
and this is related to the mass of the brane with
\be\label{mass4} M=\frac{(d-1) V_{d-1}}{16\pi G}m_0.\ee
Also, the Hawking temperature and the entropy for this solution are given by
\be\label{TS2} T= \frac{f^\prime (r)}{4 \pi}\Big|_{r=r_h}=\frac{2 d r_h^2 -  \beta^2}{8 \pi r_h}\,,\qquad S=\frac{V_{d-1}}{16\pi G} 4\pi r_h^{(d-1)}.\ee

\begin{figure}[H]
\centering
\includegraphics[width=8cm,height=6cm]{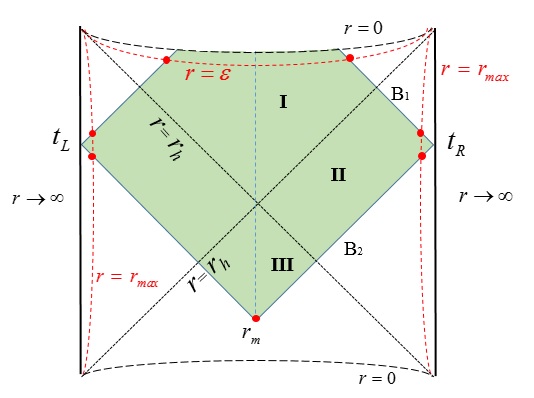}
\caption{Penrose diagram of the WDW patch of neutral AdS black brane in symmetric configuration ($t_L=t_R$).}
\label{f2}
\end{figure}

 {\bf{\emph{WDW patch:}}} The causal structure of the two-sided neutral AdS black brane with a single horizon is described by the Penrose diagram in Fig.~(\ref{f2}). The corresponding WDW patch is denoted by the shaded region bounded by the light sheets sent from the two asymptotic time slices $t_L$ and $t_R$. We choose the symmetric configuration for the time slices, i.e. $t_L=t_R\equiv t/2$. In the next subsection, we evaluate the total action (\ref{tact}) on this patch as the boundary time increases.

In this patch, $r_m$ is the point in which the past light sheets from the left and right boundaries intersect before hitting the past singularity at some critical time $t_c$ in the symmetric configuration
\be\label{critt} \frac{t_c}{2}={r}_{\infty}^{*}-r_{}^{*} (0).\ee
It also contains a cutoff surface near the future singularity at $r\!=\!\epsilon$ and two surfaces near the asymptotic boundary regions at $r\!=\!r_{max}$. These surfaces are specified by the dashed red lines in Fig.~(\ref{f2}). The boundaries $B_1$ and $B_2$ are given by Eqs.~(\ref{BB}).
%@@@@@@@@@@@@@@@@@@@
\subsubsection{The growth rate of complexity}
In the following we compute the growth rate of the holographic complexity on the WDW patch associated with a two-sided AdS black brane for times $t>t_c$ - see Fig.~(\ref{f2}).  In this regard, we consider the time dependence of the total action in (\ref{tact}). In the symmetric configuration of the WDW patch, we can perform the calculations only for the right side of the Penrose diagram or regions $\RN{1}$, $\RN{2}$, and $\RN{3}$, and then multiply the result by a factor of two.
 %$$$$$$$$$$$$$$$$$$$$$$$$$

{\bf{\emph{Bulk action:}}} The bulk contribution comes from the action (\ref{act1}) by omitting the Maxwell term. Therefore, according to the relations in (\ref{rvsv}), for $t>t_c$ we have
\bea \label{bulk2} I_{bulk}&\!\!\!\!\!=\!\!\!\!\!&2\,\, \left(I_{bulk}^{\RN{1}}+I_{bulk}^{\RN{2}}+I_{bulk}^{\RN{3}}\right)\nn\\
&\!\!\!\!\!=\!\!\!\!\!&-\frac{d\, V_{d-1}}{4\pi G}\left[\int_{\epsilon}^{r_h}\left(\frac{t}{2}+{r}_{\infty}^{*}-r_{}^{*} (r)\right)+2\int_{r_h}^{r_{max}}\left({r}_{\infty}^{*}-r_{}^{*} (r)\right)+\int_{r_m}^{r_h}\left(-\frac{t}{2}+{r}_{\infty}^{*}-r_{}^{*} (r)\right) \right]\, r^{d-1} dr,\nn\\
&\!\!\!\!\!=\!\!\!\!\!&I_{bulk}^{0}-\frac{d\, V_{d-1}}{4\pi G} \int_{\epsilon}^{r_m}\left(\frac{t}{2}+{r}_{\infty}^{*}-r_{}^{*} (r)\right) \,r^{d-1} dr,
\eea
clearly, only the second term depends on the time.
 %$$$$$$$$$$$$$$$$$$$$$$$$$

 {\bf{\emph{Boundary surface action:}}} If we choose affinely parametrization for the null normals then the null surface term vanishes ($\kappa\!=\!0$), thus we only need to consider the GHY term coming from the regulator surface at the future singularity and the surface at the UV cutoff.
 Therefore, we obtain the contribution of surface action in (\ref{surf}) for the neutral AdS solution as
 \bea\label{surf3} I_{bdy}&\!\!\!\!\!=\!\!\!\!\!&2\,(I_{bdy}^{r=\epsilon}+I_{bdy}^{r=r_{max}})\\
 &\!\!\!\!\!=\!\!\!\!\!&-\frac{d\,V_{d-1}}{8\pi G}\!\!\left[(2{r}^d-r_{h}^d)- \frac{2(d-1)r^{d-2}-dr_h^{d-2}}{2d(d-2)}\beta^2\right]\!\!\left((t/2+{r}_{\infty}^{*}-r_{}^{*} (r))\Big|_{r=\epsilon}-({r}_{\infty}^{*}-r_{}^{*} (r))\Big|_{r=r_{max}}\right).\nn
 \eea
As seen, the cutoff term at $r = r_{max}$ is independent of the time, then we can rewrite (\ref{surf3}) as
 \be\label{surf4}  I_{bdy}= I_{bdy}^0-\frac{d\,V_{d-1}}{8\pi G}\!\!\left[(2{r}^d-r_{h}^d)- \frac{2(d-1)r^{d-2}-dr_h^{d-2}}{2d(d-2)}\beta^2\right] (t/2+{r}_{\infty}^{*}-r_{}^{*} (r))\Big|_{r=\epsilon},\ee
where $I_{bdy}^0$ is independent of the time and has no contribution to the growth rate.
  %$$$$$$$$$$$$$$$$$$$$$$$$$

  {\bf{\emph{Joint action:}}} According to the WDW patch in Fig.~(\ref{f2}), there are different joint contributions at the intersection of the null boundaries with surfaces at $r=\epsilon$ and $r=r_{max}$, and with each other at $r_{m}$. However, the joints at singular and cutoff surfaces are independent of the time. Assuming that $k_1$ and $k_2$ are given by the relations in (\ref{kk}), then the joint action (\ref{joint}) can be evaluated as
 \be\label{joint3} I_{joint}=-\frac{V_{d-1}}{8\pi G} \left[r^{d-1}\,\log\frac{|f(r)|}{\xi^2}\right]_{r=r_{m}}.\ee
 The time evolution of this contribution is through the implicit time dependence of $r_m$ with equation
 \be\label{rmeq} \frac{t}{2}-{r}_{\infty}^{*}+r_{}^{*} (r_m)=0.\ee
 %$$$$$$$$$$$$$$$$$$$$$$$$$

 {\bf{\emph{Counterterm action:}}} Using the parameter expansion introduced in (\ref{amb2}), the counterterm action becomes
 \bea\label{cterm3} I_{ct}&\!\!\!\!\!=\!\!\!\!\!&2\,(I_{ct}^{future}+I_{ct}^{past}) \\
 &\!\!\!\!\!=\!\!\!\!\!& \frac{(d-1) V_{d-1}}{4\pi G}\left (\int_{{\epsilon}}^{r_{max}}+\int_{r_m}^{r_{max}}\right) r^{d-2} \log\frac{(d-1) \ell_{c}\xi}{r} dr.
 \eea
 It is clear that the cutoff bounds have no time dependence, thus in the limit ${\epsilon}\rightarrow 0$, the counterterm action depends on the time through the Eq.~(\ref{rmeq}).
%@@@@@@@@@@@@@@@@@@
\subsubsection{The late time behavior}
In the case of neutral branes we should also consider the contribution of boundary term, that is, the growth rate for $t>t_c$ is calculated from
\be\label{CAR1} \frac{d \mathcal{C}_{A}}{dt}=\frac{d}{dt} \left(\delta I_{bulk}+\delta I_{bdy}+I_{joint}+I_{ct}\right),\ee
where $\delta I_{bulk}\equiv I_{bulk}-I_{bulk}^0$ and $\delta I_{bdy}\equiv I_{bdy}-I_{bdy}^0$. From Eqs.~(\ref{bulk2}) and (\ref{surf4}) for the bulk and boundary actions,
\bea \label{joint4} \frac{dI_{joint}}{dt} &=&\frac{V_{d-1}}{8\pi G} \Bigg[r_m^d + \frac{1}{2} (d-2) r_h^d + \frac{1}{2} ( d-1) (r_m^d -  r_h^d) \log\frac{\bigl|f(r_m)\bigr|}{\xi^2} \\
	&&\hspace{1cm}+\left( \frac{( d-1) (  r_h^{ d-2}- r_m^{ d-2} ) }{4 ( d-2) }\log\frac{\bigl|f(r_m)\bigr|}{\xi^2}- \frac{1}{4} r_h^{ d-2}  \right)  \beta^2 \Bigg] \nn\eea
 for the joint term, and the following for the counterterm action
 \be\label{cterm4} \frac{dI_{ct}}{dt}=\frac{ ( d-1)\,V_{d-1}}{16\pi G} \left[ (r_m^d -  r_h^d) - \frac{( r_m^{ d-2} - r_h^{ d-2})  \beta^2}{2 ( d-2)}\right] \log\frac{( d-1) \ell_c \xi}{r_m}\,,\ee
we obtain the growth rate of the holographic complexity in the CA conjecture as
\be \label{CAR2} \frac{d \mathcal{C}_{A}}{dt}=\frac{(d-1)\,V_{d-1}}{8\pi G}\left[ r_h^d -  \frac{r_h^{d-2}  \beta^2}{2 ( d-2)}+ \tfrac{1}{2} r_m^{ d-2} f(r_m) \log\frac{( d-1)^2 \ell_c^2 |f(r_m)|}{r_m^2}\right],\ee
or equivalently, from Eq.~(\ref{mass4}) it can be recast in the form
\be\label{CAR3}
\frac{d \mathcal{C}_{A}}{dt}=2M+\frac{V_{d-1}}{16\pi G}\left[( d-1) r_m^{ d-2} f(r_m) \log\frac{( d-1)^2 \ell_c^2 |f(r_m)|}{r_m^2}\right],
\ee
where the time dependence of the meeting point $r_m$ is obtained from Eq.~(\ref{rmeq}) and definition of tortoise coordinate in Eq.~(\ref{tc}) as
\be \label{rmeq2} \frac{dr_m}{dt}=-\frac{f(r_m)}{2}.\ee

One can observe that though the result (\ref{CAR3}) does not satisfy the Lloyd's bound on the rate of quantum computation \cite{Lloyd} for $t>t_c$, but at the late time limit in which $r_m$ approaches $r_h$ and $f(r_m)\rightarrow 0$, the contribution of second term vanishes and this yields
\be\label{lbs} \frac{d\mathcal{C}_{A}}{dt}\Bigg|_{t\to \infty}= 2 M,\ee
which is consistent with the expected rate of the growth at late time in Refs. \cite{Brown:2015bva,Brown:2015lvg} even in the presence of momentum relaxation. Though the joint and the counterterm contributions of both charged and neutral solutions are sensitive to the ambiguities of null boundaries through the normalization constant $\xi$, but according to Eqs.~(\ref{CCAR2}) and (\ref{CAR2}), the result for the total action is independent of that.

More specifically, the relation (\ref{mass3}) yields a bound on the value of $\beta$ that the larger the value of $\beta$ leads to negative value for the mass which has no physical interpretation. For the neutral branes studied in this section, this maximum value is obtained from $\beta^2_{max}\!=\!2(d-2)r_h^2$. We investigate that for $\beta\leq \beta_{max}$, the equation $f(r)\!=\!0$ always has a real positive root (single horizon $r_h$) even when $m_0=0$. Further, from the definition of temperature in (\ref{TS2}) this value then yields a finite temperature $T={r_h}/{2\pi}$ in this case. In other words, the vacuum metric ($M=0$) has the form of an AdS black hole that in the context of AdS/CFT, it can be interpreted in terms of an entangled state of two copies of the
CFT on a hyperbolic plane \cite{Casini:2011kv}. The entangled state of two copies of the CFT on a hyperbolic plane appearing above can then be understood as a conformally transformed description of the global vacuum state which entangles the CFT degrees of freedom on the interior with those on the exterior of the sphere.

This result provides strong motivations to investigate the vanishing of complexity growth rate at some finite temperature other than zero. For $\beta\!=\!\beta_{max}$ this is a minimum temperature given by $T_{min}\!=\!{r_h}/{2\pi}$. Therefore, from Eq.~(\ref{lbs}) we expect that the variation of the complexity becomes zero in this temperature only  in the Lloyd limit. There is a similar discussion on the temperature for resistivity and conductivity of field theories which are dual to EMA-Dilaton theories in Ref. \cite{Gouteraux:2014hca}.

\begin{figure}[ht]
\centering
\includegraphics[width=8cm,height=6cm]{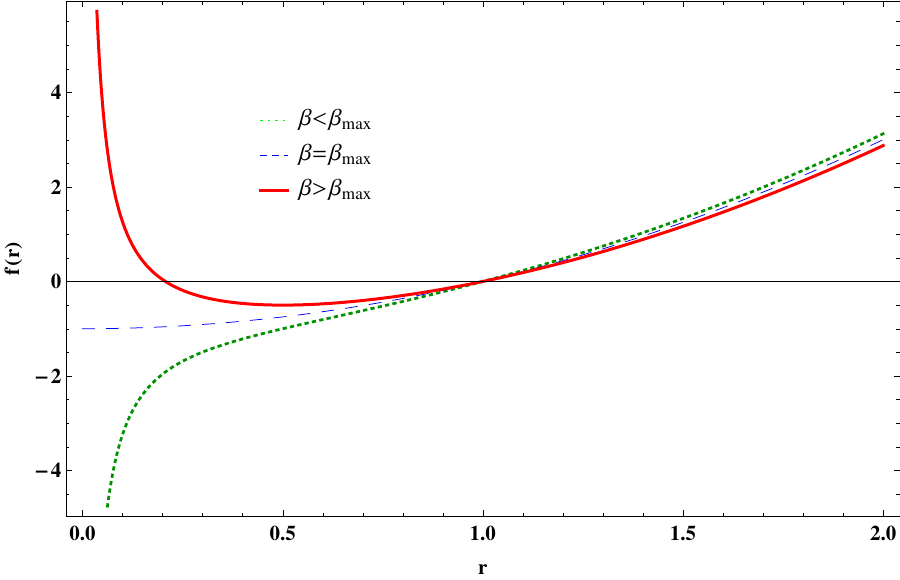}
\caption{The structure of the blackening factor for different values of $\beta$ with $d=3$ and $r_h=1$. The solid red line corresponds to $\beta>\beta_{max}$. }
\label{f3}
\end{figure}

We also note that regardless of the unphysical values obtained for the mass, in the case of $T<T_{min}$ or $\beta>\beta_{max}$, we observed that equation $f(r)=0$ has two real positive roots and the geometry have a causal structure similar to that encountered for the  charged black holes. We have plotted the behavior of the blackening factor $f(r)$ in four dimensions for different values of $\beta$ in Fig.~(\ref{f3}). The case that happens here is denoted by the solid red curve in figure. Studying the complexity growth rate for this charged-like geometry similar to what was done in the previous subsection, we find that $\dot{\mathcal{C}}_A$ vanishes at late times. It has been shown in Ref. \cite{Chapman:2016hwi} that a similar thing happens in the case of AdS black holes with hyperbolic geometry. For temperatures below $T=1/(2\pi L)$, the small hyperbolic black holes (i.e. $r_h < L$) have a causal structure similar to that of charged AdS black holes and the late time limit of $\dot{\mathcal{C}}_A$ goes to zero. In fact, though we consider the AdS branes with planar geometry in (\ref{sol1}), for $\beta=\beta_{max}$ the neutral brane metric behaves as well as a hyperbolic geometry (see appendix \ref{appA}).

\begin{figure}[H]
\centering
\subfigure[]
{\includegraphics[width=.45\textwidth]{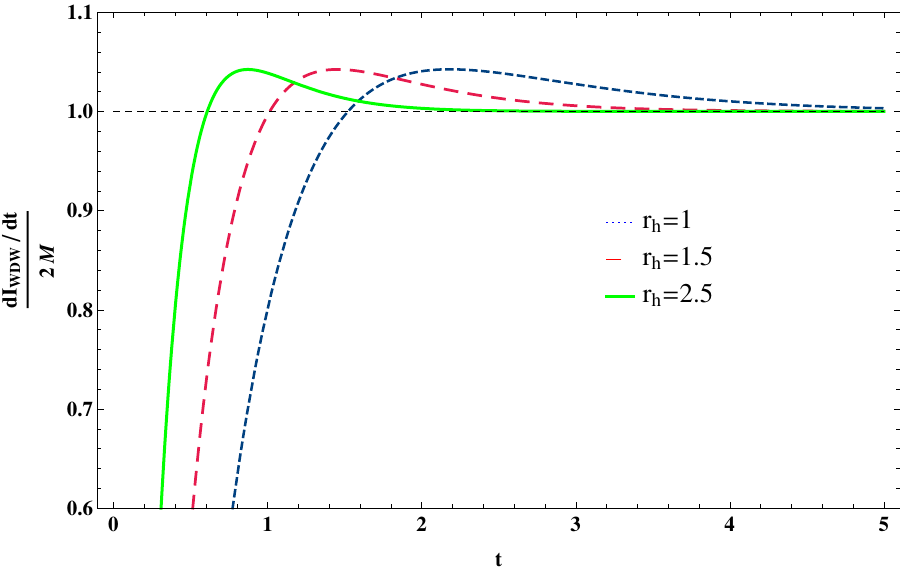}}
\hspace{5mm}
\subfigure[]
{\includegraphics[width=.45\textwidth]{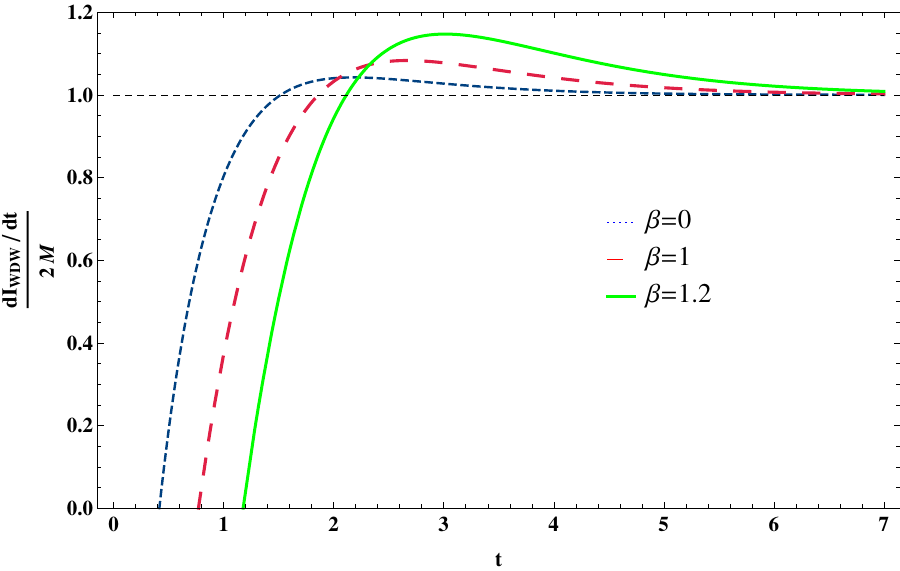}}
	    \caption{The action growth rate on the WDW patch vs. $t$ for different values of (a) $r_h$, and (b) $\beta$.}
\label{f4}
\end{figure}

%@@@@@@@@@@@@@@@@@@
To analyze the behavior of the time evolution of the holographic complexity with more detail, we investigate the time derivative of the action in Eq.~(\ref{CAR3}) as a function of time in the four-dimensional spacetime. The results are depicted for some typical locations of the event horizon, $r_h=1,1.5,2.5$, in Fig.~(\ref{f4}a) and for some values of momentum relaxation constant, $\beta=0,1,1.2$ , in Fig.~(\ref{f4}b). For the sake of the numerical precision we first solve the equation of $r_m$ in (\ref{rmeq2}) numerically and then plot the diagrams. We can see from figures that as the time passes from $t_c$, the growth rate of the action violates the Lloyd's bound for all values of $r_h$ and $\beta$, while at large times $t>>t_c$ it saturates the bound, i.e. $\dot{\mathcal{C}}_A=2M$. Although the general behavior of the plots is the same, the curves behave differently in each panel. For instance, in the left panel larger black holes saturate the bound sooner than smaller ones while all violate the bound with the same strength (the curves have the same peak) in the initial times. In the right panel, the larger value of $\beta$ corresponds to the stronger violation of bound and it saturates the bound in some later times. The other point that can be inferred from
Figs.~(\ref{f4}a) and (\ref{f4}b) is that the rate of complexity saturates the Lloyd's bound from above at late times. It seems to be consistent with the positive sign of the second term in Eq.~(\ref{CAR3}).

%%%%%%%%%%%%%%%%%%%%
\subsection{Thermal diffusivity of neutral branes at minimal temperature}
From the holographic point of view in Ref.~\cite{Brown:2015lvg}, the black holes are regarded as the fastest computers in the sense that they saturate the complexification Lloyd's bound (\ref{LB}). In this regard, the scrambling time is a measure of how long it takes for information to spread through the system of $N$ degrees of freedom \cite{Hayden:2007cs,Maldacena:2015waa}. The rate of scrambling in a chaotic system is determined by a Lyapunov time $\tau_L\sim\hbar/(2\pi k_B T)$, \cite{Maldacena:2015waa}. Quantum mechanics puts a bound on this exponent and it has been shown in Refs.~\cite{Shenker:2013pqa,Roberts:2014isa,Damle} that the black holes saturate this bound. However, a fast computer should interact strongly, so a good candidate is a strongly coupled CFT in the context of the AdS/CMT. A class of such strongly coupled theories in quantum CM physics is the notion of the strange metals with specific transport properties \cite{Hartnoll:2016apf}. Now, the question is whether the chaos properties of black holes and many body systems are connected to the transport coefficient.

It was proposed in Refs.~\cite{Hartnoll:2014lpa,Blake:2016wvh} that one can reformulate the KSS bound (bound on the ratio of shear viscosity to entropy density) \cite{Kovtun:2004de} in terms of the diffusion constant as $\mathcal{D}\sim {v_B}^2 \tau_L$ where $v_B$ is a characteristic velocity of the theory known as the butterfly velocity. For any holographic theory with a classical gravity dual both the Lyapunov time and the butterfly velocity can be extracted from properties of a black hole horizon \cite{Shenker:2013pqa,Roberts:2014isa}. On the other hand, thermal diffusivity provides a natural candidate to relate to many body chaos such that the relationship
\be\label{cp}\mathcal{D}_T\geq {v_B}^2 \tau_L,\ee
is a generic low temperature property of the homogeneous holographic lattice models \cite{Blake:2017qgd}. Indeed, it is a universal piece of the diffusivity matrix that we can generically relate to the chaos exponents at infra-red fixed points. It is defined as follows
\be\label{TD}
\mathcal{D}_T\equiv\frac{\kappa}{c_\rho},
\ee
where $\kappa$ is the open circuit thermal conductivity and $c_\rho$ is the thermodynamic specific heat at fixed density $\rho$. Though the thermal diffusion for neutral black holes in the four-dimensional EMA model with momentum relaxation has been recently studied slightly in Ref.~\cite{Jeong:2021zhz}, we consider this concept in general dimensions. On the other hand, we found that there is a minimum temperature for which the complexity growth rate vanishes just like what happens in the case of AdS black holes with hyperbolic geometry in Ref.~\cite{Chapman:2016hwi}. Therefore, it would be of interest to investigate the behavior of the transport parameters for these models at $T=T_{min}$.

In the momentum relaxation model, $T_{min}$ corresponds to $\beta_{max}$ for which the mass of the black hole becomes zero. Thus, we can recast the growth rate of the holographic complexity in Eq.~(\ref{lbs}) as
\be\label{camr}
\dot{\mathcal{C}}_A=2 M=\frac{2 (d-1)  }{d-2} \, S \,(T-T_{min}),
\ee
where $S$ and $T$ are the entropy and the temperature of neutral AdS black branes given in Eqs.~(\ref{TS2}). Now following \cite{Blake:2017qgd}, we can calculate the thermal conductivity and the specific heat respectively from
\be\label{kcn1}
   \kappa = 4 \pi \,\frac{ f'(r) r^{2(d - 2)} }{ (f'(r) r^{d - 2})'}\bigg |_{r=r_h}, \qquad c_{\rho} = T\frac{\partial s}{\partial T},
\ee
where $s={S}/{V_{d-1}}$ is the entropy density and we set $16\pi G\!=\!1$ in the rest of this section and Appendix \ref{appA}.
Substituting the blackening factor (\ref{sol1}) with $q=0$ and (\ref{TS2}) into Eqs. (\ref{kcn1}) then yields
\be\label{kcn2}
\kappa = 4 \pi r_h^{ d-2}\,\frac{ (-2 d r_h^2 + \beta^2)}{2 d ( d-3) r_h^2-( d-1) \beta^2} , \qquad c_{\rho} ={4\pi( d-1) r_h^{ d-1}}\, \frac{  (2 d r_h^2 -  \beta^2)}{ (2 d r_h^2 + \beta^2)}.
\ee
Now, if one inserts these parameters in Eq. (\ref{TD}) the diffusion constant is given by
\be\label{tdn2}
   \mathcal{D}_{T} =\frac{1 }{( d-1) r_h } \, \frac{ (2 d r_h^2 + \beta^2)}{  \bigl(( d-1) \beta^2 -2 d ( d-3) r_h^2  \bigr)} \,.
\ee
For the minimum value of the temperature where $\beta^2_{max}=2 ( d-2) r_h^2$, the above transport parameters are reduced to
\be\label{kcdnmin}
   \kappa =4 \pi r_h^{ d-2} , \qquad c_{\rho} =4 \pi r_h^{ d-1} , \qquad \mathcal{D}_{T} = \frac{ 1}{ r_h} \,.
\ee
 The calculations for the AdS hyperbolic black holes are left to Appendix \ref{appA}. From the relations (\ref{kcdnmin}) and (\ref{kcdhmin}), one can find that irrespective of the dimension of spacetime, we have the same expressions for these quantities in both solutions at the minimum temperature. In particular, the diffusion constants are independent of the spacetime dimensions and are proportional to the inverse of the horizon radius.  Such effect has long been similarly established in Ref. \cite{Iqbal:2008by} that the DC transport coefficients of conserved quantities is related to the horizon via the membrane paradigm.

As previously mentioned, there is a deep connection between the transport coefficients and the chaos parameters of any holographic geometry. Moreover, these parameters can be extracted by analyzing a shock wave propagating on the black hole horizon\cite{Shenker:2013pqa,Roberts:2014isa,Blake:2016wvh}. In units $\hbar\!=\!k_B\!=\!1$, the Lyapunov time is given by $\tau_L\sim (2\pi T)^{-1}$ and the butterfly velocity can be calculated in the near horizon of isotropic metric (\ref{ansatz1}) like \cite{Baggioli:2018afg,Feng:2017wvc}
 \be\label{bv1}
 v_{B}^2=\frac{ f(r)'  }{2 ( d-1)  r}\bigg |_{r_h}=\frac{2 d r_h^2 -   \beta^2}{4 ( d-1) r_h^2},
 \ee
 where we have used the blackening factor (\ref{sol1}) with $q=0$ for neutral black branes. In the minimum temperature or equivalently $\beta^2_{max}=2 ( d-2) r_h^2$, this velocity becomes
 \be\label{bv2}
v_{B}=\sqrt{\frac{1}{d-1}}=\sqrt{\frac{2}{d}}\,{v_{B}^{Sch}},
\ee
 where $v_{B}^{Sch}$ is the value of the butterfly velocity for an AdS-Schwarzschild black brane in $d+1$ dimensions \cite{Shenker:2013pqa}. As shown the velocity in EMA theory at this special point is lower than its counterpart in Einstein gravity for $d\geq 3$. The diffusion constant in (\ref{kcdnmin}) together with parameters $v_B$ and $\tau_L$ at the minimal temperature $T_{min}=r_h/2\pi$ respect the bound in (\ref{cp}), i.e.,
 \be\label{db}\frac{D_{T}}{\tau_{L} \,v_{B}^2}= (d-1)\geq 1.\ee
 As is obvious they saturate the bound only for $d=2$ in the three-dimensional spacetime.
%%@@@@@@@@@@@@@@@@@@@@@@@@@@@
\section{The growth rate of complexity in the k-essence sector }
The previous studies can be generalized to the case in which the kinetic term for the scalar fields can have non-linear contributions. Such a case can be implemented by the
so-called k-essence models \cite{ArmendarizPicon:1999rj}, in which the mentioned kinetic term are generalized to be a function $P( \psi,(\prt\psi)^2)$. A simple case contained in this setup is that the scalar fields apart from the standard kinetic term possess a kinetic non-linear contribution given by the higher powers of the kinetic term. In this subsection, we are going to study the growth rate of complexity for the dyonic AdS black branes in this holographic model from the CA proposal. However, for later convenience we consider the four-dimensional bulk spacetime.

The action of the non-linear EMA theory was studied in Refs. \cite{Baggioli:2014roa,Cisterna:2017jmv} and is given by
\be\label{nlact} I_{bulk}=\frac{1}{16\pi G}\int_{M} d^{4}x \,\sqrt{-g}\left[R-2\Lambda-\frac14 F_{\mu\nu}F^{\mu\nu}- \sum_{I=1}^{2} (\chi_I+ \gamma \chi_I^k)\right], \ee
where $\chi_I=\frac12 \prt_{\mu} \psi_{I} \prt^{\mu}\psi_{I}$ and $\gamma$ is the coupling of non-linear axionic term. The AdS black brane solution of this model is described by the ansatz (\ref{ansatz1}). Therefore, the blackening factor and the axionic scalar fields are
\be\label{sol2} f(r) = r^2 - \frac{\beta^2}{2}  + \frac{Q_e^2+Q_m^2}{4r^2} - \frac{2 m_0}{r} +\gamma \,\frac{ \beta^{2 k}}{2^{k} (2 k-3) r^{2(k-1)} },\quad
 \psi_1=\beta x_1,\,\,\psi_2=\beta x_2\,.\ee
The Maxwell equations is easily solved by
\be\label{GF}A=-\frac{Q_e}{r}\,dt+\frac{Q_m}{2}\left(x_1 dx_2-x_2 dx_1\right).\ee
where $Q_e$ and $Q_m$ are the electric and magnetic monopole charges.
The temperature and the entropy of solution are given by
\be\label{mc}  T\!=\!\! \frac{f^\prime (r_+)}{4 \pi}\!=\!\frac{1}{4\pi}\!\left(3r_+  -\frac{\beta^2}{2r_+}-\frac{Q_e^2+Q_m^2}{4r_+^3}-\gamma \frac{\beta^{2k}}{2^k  r_+^{2k-1}}\right)\!\!,\quad S\!=\!\frac{V_2}{16\pi G} 4\pi r_+^{2}.\ee
Also, the mass parameter is obtained from condition $f(r_+)=0$ such that the mass of solution becomes
\be \label{mass5} M\!=\!\frac{4V_2}{16\pi G}m_0,\qquad m_0 = \frac{Q_e^2+Q_m^2}{8 r_+} + \frac{1}{2} r_+^3 -  \frac{1}{4} r_+ \beta^2 -\gamma \frac{\beta^{2 k} }{2^{k+1}  (2 k-3)  r_+^{ 2 k-3} },\ee
and $r_+$ is the location of the event horizon. The extended thermodynamics of the above solution has been studied in Refs. \cite{Cisterna:2017jmv,Hu:2018prt}. We note also that in particular, for $k=2$ the metric (\ref{sol2}) behaves as a double horizons black hole, i.e. $r_+$ and $r_-$, so one can use the WDW patch in Fig.~(\ref{f1}) to compute the evolution of the holographic complexity of state which is dual to this geometry.
%@@@@@@@@@@@@@@@@@@
\subsection{The action growth rate}
The total time derivative of the holographic complexity for dyonic charged AdS solutions in this model is calculated from (\ref{CCAR1}). However, as mentioned in the previous section, we can also consider the contribution of a boundary term for the Maxwell field in the total action as \cite{Goto:2018iay}
\be\label{mq} I_{\mu Q}=\eta \int_{\prt \mathcal{M}} d\Sigma_{\mu} F^{\mu\nu} A_{\nu},\ee
which does not change the equations of motion. In general, employing a Dirichlet boundary condition results in a well-posed variational principle, but due to the boundary term (\ref{mq}) we instead need to impose a Neumann boundary condition for $\eta=1$, or a mixed boundary condition for general $\eta$. A comprehensive discussion about this boundary action is given in Ref. \cite{Goto:2018iay}. On the other hand, using the Stokes's theorem and the Maxwell equations we can convert the boundary term (\ref{mq}) to the bulk Maxwell action as
\be\label{mq1} I_{\mu Q}=\frac{\eta}{2} \int_{\mathcal{M}} d^4 x \sqrt{-g} F_{\mu\nu} F^{\mu\nu} ,\ee
therefore, the boundary action (\ref{mq}) contributes in the complexity growth rate through this bulk term.

Using similar discussion for the three bulk regions on the WDW patch in Fig.~(\ref{f1}), the complexity of the bulk action is written as follows
\be \label{bulk3} I_{\text{bulk}}+I_{\mu Q}=\frac{V_2}{16\pi G} \int_{WDW}\!\!\! dr\,dt \, r^{2} \Bigg[ -6 - \gamma \frac{\beta^4 }{2 r^4} -  ( 2 \eta-1) \frac{Q_e^2 -  Q_m^2}{2 r^4}\Bigg],\ee
where in four dimensions the volume of the boundary surface becomes $V_2$, then for the time evolution of the bulk term we have
\be\label{NCAR1} \frac{d }{dt }(\delta I_{\text{Bulk}}+I_{\mu Q})=  \frac{V_2}{8 \pi G } \Bigg[  r^3- \gamma\frac{ \beta^4 }{4 r} - ( 2 \eta-1 )  \frac{Q_e^2 -  Q_m^2}{4 r}  \Bigg]{\Bigg|}_{r_m^2}^{r_m^1}\,.\ee
For the joints of null boundaries in $r_{m}^1$ and $r_{m}^2$ we have
\be\label{NCAR2}  \frac{dI_{joint}}{dt}=- \frac{V_2}{8\pi G} \left[r^3-\frac{Q_e^2+Q_m^2}{4r}-\gamma \frac{\beta^4}{4\,r}+r f(r) \log\frac{|f(r)|}{\xi^2}\right]{\Bigg|}_{r_m^2}^{r_m^1},\ee
and also the contribution of the counterterm action becomes
\be\label{NCAR3} \frac{dI_{ct}}{dt}= \frac{V_2}{8\pi G} \left[4 m_0 -  \frac{Q_e^2 + Q_m^2}{2 r} -  \frac{ \beta^4 \gamma}{2 r} + \beta^2 r - 2 r^3  \right]\log\frac{2 \ell_c \xi}{r}  {\Bigg|}_{r_m^2}^{r_m^1}.\ee

As asserted in the previous section, the contribution of the GHY boundary terms for cutoff surfaces on the WDW patch in Fig.~(\ref{f1}) is time independent, thus we ignore them.
Combining the above results we can determine the final expression for the complexity growth rate as
\be\label{NCAR4} \frac{d \mathcal{C}_{A}}{dt}=\frac{V_2}{ 8\pi G }
\left[\frac{  (1 -  \eta)\,Q_e^2 +  \eta\, Q_m^2}{2r}- r f(r) \log\frac{4 \ell_c^2 |f(r)|}{r^2}\right]_{r_m^2}^{r_m^1}.\ee
It is obvious that the normalization constant $\xi$ and the coupling of the non-linear axionic action $\gamma$ do not affect the final result explicitly. Also, one can check that in the limit $\gamma\to 0$, Eqs.~(\ref{NCAR1})-(\ref{NCAR3}) give the results in the previous subsection for the electrically ($Q_m=0$) charged AdS$_{d+1}$ black branes when $d=3$. According to Eq.~(\ref{NCAR1}), for $\eta=1/2$ the boundary and the bulk Maxwell terms cancel each other and do not affect the gravitational action, in spite of their geometric contribution in the background (\ref{sol2}) to the action.
%@@@@@@@@@@@@@@@@@@@@@@@
\subsection{The late time behavior}
 Since in the late time limit the meeting points reach to the horizons of the geometry on the WDW patch, as shown in Fig.~(\ref{f1}), the final result for the growth rate of complexity is
\be\label{NCAR5} \frac{d\mathcal{C}_{A}}{dt}\Bigg|_{t>>t_c}=\frac{V_2}{16\pi G}\Big[\frac{  (1 -  \eta)\,Q_e^2 +  \eta\, Q_m^2}{r}\Big]_{r_+}^{r_-}.\ee
The result shows that in the absence of Maxwell boundary term, i.e. $\eta=0$, the late time behavior is similar to the case of charged black branes in (\ref{CCAR3}) even in the presence of the magnetic charge. Also for $\eta=1$, the late-time growth rate is only proportional to the magnetic charge, i.e. it vanishes for electrically charged black branes.
%@@@@@@@@@@@@@@@@@@@@@@@

\begin{figure}[H]
\centering
\subfigure[$Q_e/Q_m=2$]
{\includegraphics[width=.48\textwidth]{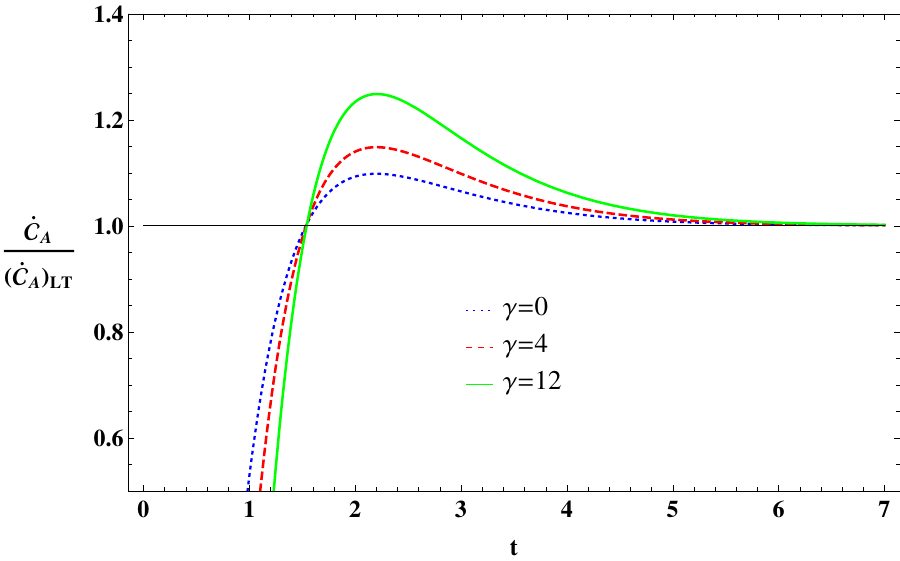}}
\hspace{3mm}
\subfigure[$\gamma=4$]
{\includegraphics[width=.48\textwidth]{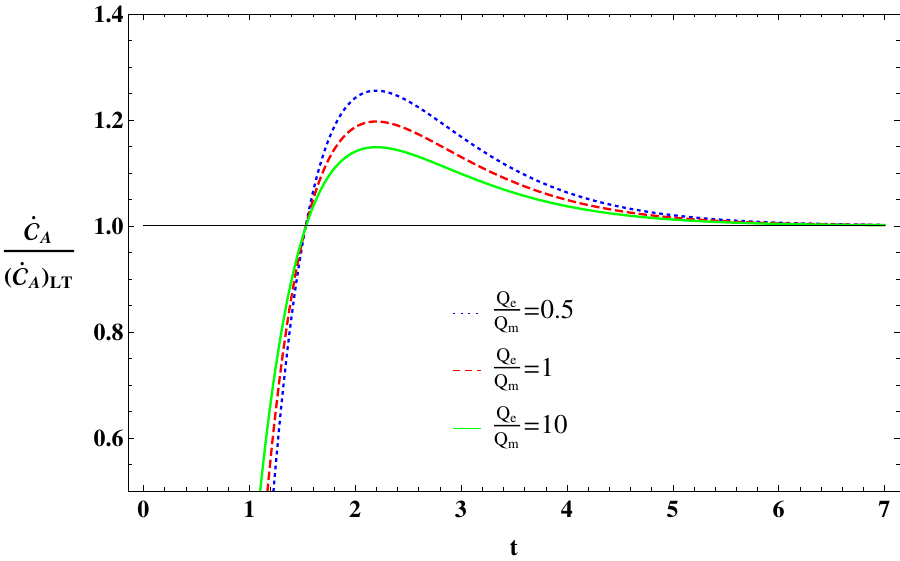}}
	    \caption{Lloyd's bound violation for different values of $\gamma$ and $Q_e/Q_m$  with $\beta=0.2, \eta=0.2, \ell_c=1$.}
\label{f5}
\end{figure}

\begin{figure}[H]
\centering
\includegraphics[width=8cm,height=5cm]{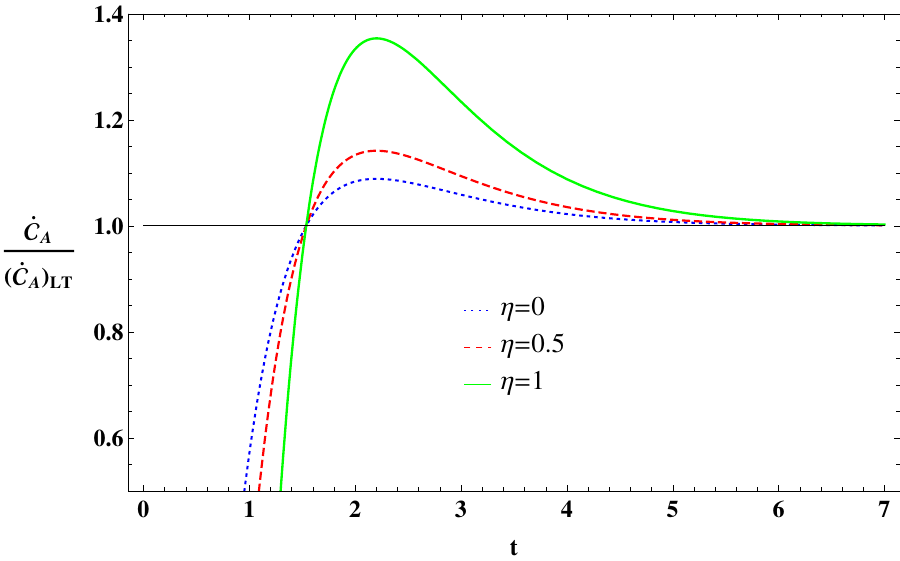}
\caption{Lloyd's bound violation for different values of $\eta$ with $Q_e/Q_m=2, \beta=0.2, \gamma=5, \ell_c=1$.}
\label{f6}
\end{figure}

To better understanding the full time-dependence of Eq.~(\ref{NCAR4}), it is straightforward to provide a numerical study on the growth rate of complexity for the dyonic black holes described by (\ref{sol2}). In this respect, we have plotted the ratio $\dot{\mathcal C}_A/(\dot{\mathcal C}_A)_{LT}$ for different values of $\gamma$, $Q_e/Q_{m}$, and $\eta$ in Figs.~(\ref{f5}) and (\ref{f6}). Here, ``LT'' stands for the late time behavior and $(\dot{\mathcal C}_A)_{LT}$ is given in Eq.~(\ref{NCAR5}). The figures show the bound is violated due to the fact that $\dot{\mathcal C}_A$ approaches the bound at very late times from above.

We see from Fig.~(\ref{f5}a) that irrespective of the behavior of $\dot{\mathcal C}_A$ at the initial times, the larger the value of non-linear axionic term, the stronger violation of the bound. In contrast, by increasing the ratio between the electric and the magnetic charges, the violation becomes weaker as shown in Fig.~(\ref{f5}b). This opposite behavior relative to these constant parameters is expectable due to the relations (\ref{sol2}) and (\ref{NCAR4}). In other words, for dominant electrical solution, i.e. $\eta<1/2$, the sign of charges ratio and $\gamma$ term are different. Of course, we have checked that this opposition also occurs for dominant magnetic solution ($\eta>1/2$). We have checked that the behavior of $\dot{\mathcal C}_A/(\dot{\mathcal C}_A)_{LT}$ for charged black branes, obtained in (\ref{CCAR2}) via CA conjecture, is very similar to the dotted plot shown in Fig.~(\ref{f5}a) with $\gamma=0$ and $Q_m=0$.
In Fig.~(\ref{f6}) we illustrate the effect of the boundary Maxwell action (\ref{mq}) with coupling $\eta$. As observed, when one increases the value of $\eta$ the Lloyd's bound for these charged solutions is violated from above drastically, just like what happens for the values of the non-linear axionic term in Fig.~(\ref{f5}a).

 %@@@@@@@@@@@@@@@@@@@@@@
\section{Conclusions and outlook}
In this paper, we extended the study of holographic complexity via AdS black branes with momentum relaxation in $(d+1)$-dimensional EMA theory by using the CA conjecture. The momentum relaxation introduced by linear massless axion fields in the bulk breaks the translational symmetry of the dual field theory and gives finite conductivity. In this regard, to retain the homogeneity of the bulk theory, the axion fields have been assumed to be a linear function of the boundary spatial coordinates with the equal constant coefficients determining the strength of momentum relaxation. Particularly, we have investigated the effects of this parameter on the holographic complexity and its time evolution.

We computed the complexity for both the charged and neutral branes as homogeneous and isotropic solutions of the bulk theory. Following the approach in \cite{Carmi:2017jqz} to evaluate the complexity on the WDW patch in each sector, we considered the general GHY surface terms and the contribution of joints and counterterm of the corresponding null boundaries. The late time behavior of the growth rate of the holographic complexity was studied for two solutions and we found that it is always finite and well defined, and satisfies the Lloyd's bound in Eqs.~(\ref{CCAR3}) and (\ref{lbs}). It has been found in Ref. \cite{Brown:2015bva} that the action growth rate at late times for neutral AdS black holes is $d I_{WDW}/dt=2M$, independent of the size of the black hole and the spacetime dimension. We observed that the neutral AdS black brane confirms this statement even in the presence of momentum relaxation.

We also obtained an upper bound for the strength of the momentum relaxation $\beta_{max}$, in the case of neutral branes, such that for $\beta>\beta_{max}$ the brane mass becomes negative which has no physical meaning. We studied the time evolution of the complexity numerically in Figs.~(\ref{f3}) for different values of $r_h$ and $\beta$. The plots showed that in spite of different rates, the complexity growth rates for different sizes of the brane and different strengths of the momentum relaxation saturate the Lloyd's bound from above. An outstanding outcome corresponding to $\beta_{max}$ is that there is a minimum temperature given by $T_{min}={r_h}/{2\pi}$ for which the growth rate of complexity vanishes in spite of the fact that it should be vanished at zero temperature for neutral AdS branes.
In a separate development, remarkable connections have been pointed out between
the dynamics of black holes and the nature of quantum chaos in many-body quantum
systems in the context of the holographic correspondence. We computed the thermal conductivity and diffusivity for neutral AdS black branes and hyperbolic black holes in general $d+1$ dimensions as well. We have shown that at the minimum temperature, the diffusion constants are proportional to the inverse of the event horizon radius independent of the dimension of spacetime. Also, this coefficient accompanied with the chaotic parameters respected the corresponding bound in the CM physics and saturate this bound at minimum temperature only in the case of three dimensions.

In addition, we have studied a holographic model including non-linear contribution of axionic kinetic term while preserves the homogeneity and isotropy of the solutions. We assumed a particular branch of solutions that has been described by a dyonic charged black brane with momentum relaxation and then calculated the rate of complexity in this model. We have also considered a boundary action for the Maxwell field in this model. The Maxwell boundary term contributed as a bulk action in the change of complexity such that for $\eta=1/2$ the bulk actions had no contribution in the complexity rate.
The results showed that even though the coupling of the non-linear term ($\gamma$) affects the contribution of different actions in the total complexity, it does not change the growth rate at late times as denoted in Eq.~(\ref{NCAR5}).

In the absence of Maxwell surface term ($\eta=0$), the growth rate vanished at late time for purely magnetic charged branes, while for non-zero electric charges it gave the known expression for general charged branes as in (\ref{CCAR3}). In contrast, for $\eta=1$ the behavior is reversed, that is the rate of growth is non-vanishing for pure magnetically charged branes and vanishes for electrically ones. The numerical investigation for the full time dependence of the complexity in the CA proposal illustrated in Figs.~(\ref{f5}) and (\ref{f6}). The results show that the Lloyd's bound is violated due to the fact that $\dot{\mathcal C}_A$ approaches the bound at very late times from above even for different values of $\gamma$, $Q_e/Q_{m}$, and $\eta$, of course with different rates.

It would be of interest to consider the effects of momentum relaxation on the growth rate of complexity in the case of charged dilatonic backgrounds. In general, the action (\ref{act1}) in the presence of dilaton and axion fields with momentum relaxation recasts as follows \cite{Gouteraux:2014hca}
 \be\label{dil}  I=\frac{1}{16\pi G}\int_{M} d^{d+1}x \,\sqrt{-g}\left[R-\frac12 (\prt \phi)^2+V(\phi)-\frac14 Z(\phi) F_{\mu\nu}F^{\mu\nu}-\frac12 \sum_{j}^{d-1} (\prt \psi_{j})^2\right].\ee
The related discussions about the growth of the holographic complexity for dilatonic metrics without axion fields have been done in Ref. \cite{Swingle:2017zcd}. For instance, it has been shown in Ref. \cite{Cai:2017sjv} that the total rate of the holographic complexity at late times is given by
\be\label{CRD} \frac{d \mathcal{C}_{A}}{dt}\Bigg|_{t\to \infty}=2M-\mu Q-D, \qquad \qquad D\equiv \frac{e^{2\phi}Q^2}{2M}.\ee
We suggest that adding the axionic action, as in (\ref{dil}), will change the structure of the Lloyd's bound of complexity given in (\ref{CRD}) by a term like $D$, however we postpone the study of this proposal for future works.

One can also study the complexity growth rate of a non-relativistic but isotropic boundary theory \cite{Ling:2016ien} which
is dual to a bulk geometry with momentum relaxation for the Lifshitz and hyperscaling violating metrics, as done in Ref. \cite{Alishahiha:2018tep} for EMD theory without momentum relaxation. Another proposal in the context of CA conjecture for these holographic models is to investigate the complexity growth rate of AdS black branes at a finite cutoff. The concept of this geometric cutoff at $r = r_c$ comes from the $T\bar{T}$ deformation of a conformal field theory in the AdS/CFT dictionary \cite{Zamolodchikov:2004ce,McGough:2016lol}, such that the coupling of this operator removes the asymptotic region of the AdS spacetime. In this regard, there have been made some efforts in Refs. \cite{Akhavan:2018wla,Alishahiha:2019lng,Hashemi:2019xeq}.
%%%%%%%%%%%%%%%%%%%%%%%%%%%%%%%%%
\appendix
\section{Thermal diffusivity of hyperbolic black holes}
\label{appA}
In this appendix we consider the thermodynamics and transport properties of neutral AdS black holes in $d+1$ dimensions with hyperbolic geometry. Following the convention in \cite{Chapman:2016hwi}, the metric with spherical symmetry takes the general form
\be \label{HM}
d s^{2} = - f(r)\, d t^{2} + \frac{d r^{2}}{f(r)} + r^{2}\, d \Sigma^{2}_{k,d-1}\,,
\ee
where the blackening factor is given by
\be\label{mf}
f(r) = \frac{r^2}{L^2}+k -\frac{\omega^{d-2}}{r^{d-2}}\,,
\ee
in which $\omega$ is the mass parameter and $L$ denotes the AdS curvature scale. $d\Sigma^2_{k,d-1}$ is the ($d$-1)-dimensional line element of curvature $k=\lbrace+1,0,-1\rbrace$ so that the black holes corresponding to $k=\{+1, 0, -1\}$ have spherical, planar, and hyperbolic horizons, respectively.

Here, we are interested in case $k=-1$ where $d\Sigma^2_{-1,d-1}=d\theta^2+\sinh^2\theta\, d\Omega^2_{d-2}$ is the metric on a ($d-1$)-dimensional hyperbolic plane. The mass of the black hole is given by
\be\label{mass}
M = \frac{(d-1) \, V_{d-1}}{16 \pi \, G}\, r_h^{d-2}\left(\frac{r_h^2}{L^2}-1\right),
\ee
where $V_{d-1}$ denotes the dimensionless volume of the relevant spatial geometry and $r_h$ is the event horizon of the black hole whose position is the largest root of $f(r_h)=0$.

 The entropy and Hawking temperature of the black hole are
\be\label{sthbh}
S =\frac{A_h}{4G}=\frac{V_{d-1}}{4 G}\,r_h^{d-1},\qquad
T=\left.\frac{f'(r)}{4\pi }\right|_{r=r_h}=\frac{1}{4\pi \,r_h}\left(d\,\frac{r_h^2}{L^2} - (d-2)\, \right).
\ee
The minimum temperature for which the rate of growing the holographic complexity vanishes is given by $T_{min}=\frac{1}{2\pi L}$. Therefore, one can rewrite its rate at late time limit as
\be\label{cah}\dot{\mathcal{C}}_A=2M=\frac{2(d-1)}{d} S\,(T-T_{min}).\ee

Now, we consider the thermal conductivity and diffusivity for hyperbolic black holes in this minimal temperature. From the definitions in Eq.~(\ref{kcn1}) and the entropy and temperature in Eq.~(\ref{sthbh}) we have
\be\label{kch}
   \kappa =4 \pi r_h^{ d-2}\,\frac{\bigl(d r_h^2-( d-2) L^2 \bigr)}{d ( d-3) r_h^2-(d-1)(d-2) L^2}, \qquad c_{\rho} =4 \pi ( d-1) r_h^{ d-1} \frac{\bigl(  d r_h^2-(  d-2) L^2\bigr) }{ \bigl(  d r_h^2+( d-2) L^2\bigr)}\,.
\ee
Also, Eq.~(\ref{TD}) yields the diffusion constant
\be \label{tdh}
 \mathcal{D}_{T} =-\frac{1}{( d-1) r_h}\frac{d r_h^2+( d-2) L^2}{ \bigl(d( d-3) r_h^2-(d-1)(d-2) L^2\bigr)}.
\ee
Finally we obtain the following expressions in the minimum temperature for which $L=r_h$,
\be\label{kcdhmin}
   \kappa =4 \pi r_h^{ d-2}, \qquad c_{\rho} =4 \pi\,r_h^{ d-1},  \qquad \mathcal{D}_{T} =\frac{ 1}{ r_h}.
\ee

%%%%%%%%%%%%%%%%%%%%%%%%%%%%%%%%%
\section*{Acknowledgment}
The authors would like to thank M. Alishahiha, A. Ghodsi, M. R. Mohammadi Mozaffar and G. Jafari for valuable comments and discussions. HBA and HM would like to thank Z. Ebadi, M. Nattagh Najafi and M. Maleki for useful discussions.
%@@@@@@@@@@@@@@@@@@@@@@@@@@@@@@@@@@
%%%%%%%%%%%%%%%%%%%%%%%%%%%%%

   \end{document}